\documentclass[aps,prl,twocolumn,showpacs,superscriptaddress]{revtex4}
\usepackage{graphicx}
\usepackage{amsmath}
\usepackage{amssymb}

\input{epsf}
\begin{document}

\title{Generalized Efimov scenario for heavy-light mixtures}

\author{D. Blume and Yangqian Yan}
\affiliation{Department of Physics and Astronomy,
Washington State University,
  Pullman, Washington 99164-2814, USA}

\date{\today}

\begin{abstract}
Motivated by recent experimental investigations
of Cs-Cs-Li Efimov resonances,
this work theoretically investigates the few-body properties
of $N-1$ non-interacting identical heavy bosons, which
interact with a light impurity through a large $s$-wave
scattering length.
For Cs-Cs-Cs-Li, we predict the existence of universal four-body
states with energies
$E_4^{(n,1)}$ and $E_4^{(n,2)}$, which are universally linked to
the energy $E_3^{(n)}$ of the $n$th Efimov trimer.
For infinitely 
large $^{133}$Cs-$^6$Li 
and vanishing $^{133}$Cs-$^{133}$Cs scattering lengths,
we find $(E_4^{(1,1)}/E_3^{(1)})^{1/2} \approx 1.51$ and
$(E_4^{(1,2)}/E_3^{(1)})^{1/2} \approx 1.01$.
The $^{133}$Cs-$^6$Li 
scattering lengths at which  these states merge with the
four-atom threshold, the dependence of these
energy ratios on the mass ratio between the heavy and light atoms, and
selected aspects of the generalized Efimov scenario for $N>4$ are 
also discussed.
Possible implications of our results for ongoing cold atom experiments
are presented.
\end{abstract}


\maketitle
Continuous and discrete scale invariances underlie many phenomena in
physics. The possibly most aesthetically appealing
examples are fractals~\cite{Mandelbrot}, where a given pattern repeats
itself as one zooms in.
Scale invariance phenomena also emerge in quantum mechanics.
A prominent example is the three-body Efimov 
effect~\cite{Efimov,BraatenReview}.
If there exists an Efimov trimer of size
$l_3^{(n)}$ and with energy $E_3^{(n)}$~\cite{energyscale}, 
then there should exist
another larger and less strongly-bound Efimov trimer 
of size $l_3^{(n+1)}=\lambda l_3^{(n)}$
and with energy $E_3^{(n+1)} = \lambda^{-2} E_3^{(n)}$.
Here, $\lambda$ ($\lambda>1$) is a scaling factor that depends 
on the masses and particle statistics of the constituents.

The experimental observation of consecutive three-body resonances
is extremely challenging as it requires working in the
universal Efimov window.
To be in this window, the absolute values of at least two of
the three two-body $s$-wave scattering lengths~\cite{footnote2vs3} 
have to be larger
than the other length scales of the underlying two-body potentials
and the temperature has to be lower than the energy scale set
by the $s$-wave scattering length. Thus, to observe two consecutive 
three-atom resonances, exquisite control
over the scattering lengths and ultralow temperatures are required.
For three identical bosons, $\lambda$ is approximately
equal to $22.7$ and two consecutive three-atom resonances 
in a bosonic system have only been observed recently in 
$^{133}$Cs~\cite{Grimm2014,footnoteresonances}.

It is well known 
that the scaling factor $\lambda$ takes smaller, and hence 
more favorable, values for
heteronuclear mixtures with infinitely large
interspecies $s$-wave 
scattering length~\cite{BraatenReview,Efimov2,Efimov3,petrov,esry2,esry1,hammer2}.
For 
$^{133}$Cs-$^{133}$Cs-$^6$Li, e.g., $\lambda$ takes the 
value 
$4.877$.
For notational convenience, we use Cs and Li to refer to the bosonic
$^{133}$Cs and fermionic $^{6}$Li isotopes in what follows.
Indeed, recently the Heidelberg~\cite{Heidelberg} and Chicago~\cite{Chicago} 
groups independently reported
the experimental observation of, respectively, 
two and three consecutive Cs-Cs-Li three-atom
resonances.
The analysis shows that the Cs-Cs interactions play a 
negligible role
at the present precision of the experiments, indicating
that
the observation of Efimov physics in these heavy-light mixtures
is due to the large magnitude of the Cs-Li $s$-wave scattering 
length.

The extended Efimov scenario has been studied 
predominantly for four {\em{identical}} bosons with large
in absolute value
two-body $s$-wave scattering 
length~\cite{Platter,vonStecher,Deltuva1,Deltuva,potassiumspacing,Hulet,Grimm4body}.
In this case, there exist two four-body
states with energies $E_4^{(n,1)}$ and $E_4^{(n,2)}$ that
are universally tied to the $n$th Efimov trimer with
energy $E_3^{(n)}$.
These four-body states lead to measurable four-atom
resonances on the negative scattering length side
(at 
scattering lengths $a_{4,-}^{(n,1)}$ and $a_{4,-}^{(n,2)}$)
and atom-trimer and dimer-dimer resonances on the
positive scattering length side~\cite{potassiumspacing,Hulet,Grimm4body}.

This Letter explores the generalized Efimov scenario for
$N-1$ identical heavy bosons and a single light impurity
for the case where the magnitude of the heavy-light $s$-wave scattering length
is large
compared to all other two-body length scales, including the
heavy-heavy $s$-wave scattering length.
For the Cs-Cs-Cs-Li system,
we find---as was found for four
identical bosons---two tetramer states at unitarity.
Moreover,
we find that these four-body states become unbound at
Cs-Li $s$-wave scattering lengths
$a_{4,-}^{(1,1)} \approx 0.55 a_{3,-}^{(1)}$ and 
$a_{4,-}^{(1,2)} \approx 0.91 a_{3,-}^{(1)}$.
Since
$a_{4,-}^{(1,2)}$ is close to
$a_{3,-}^{(1)}$, the
loss features in the Heidelberg and Chicago 
experiments~\cite{Heidelberg,Chicago} that were identified as being due to
three-body physics could potentially, in the proper temperature 
and density regime,
include a ``contamination'' from the four-body sector.
Our calculations thus suggest
that it would be extremely interesting to search for universal
four-body physics in Cs-Li mixtures.
When the mass ratio $\kappa$
between the heavy and light atoms  
is reduced to less than $\approx 13$,
the energy of 
the excited tetramer at unitarity lies above that of the trimer.
For very large mass ratios, we find---as
in the case of the Cs-Cs-Cs-Li 
system---two tetramers at unitarity.

An intriguing question is how the extended Efimov scenario, if
existent, looks for $N>4$. For
$N$ identical bosons,
evidence has been presented that there exist five-body
and higher-body states that are
universally tied to each Efimov 
trimer~\cite{vonStecherJPB,vonStecherPRL,Kievsky}.
While many questions
regarding the $N>4$ extension of the Efimov scenario for identical
bosons remain~\cite{hanna,vonStecherJPB,Nicholson,Kievsky2},
essentially nothing is known about heteronuclear systems 
with $N>4$.
We find five- and six-body states for the B$_{N-1}$X system
that are universally tied to the lowest Efimov trimer.

Our model Hamiltonian $H$,
\begin{eqnarray}
\label{eq_ham}
H = -\frac{\hbar^2}{2m_{\text{B}}} \sum_{j=1}^{N-1} \nabla^2_{\vec{r}_j}
-\frac{\hbar^2}{2m_{\text{X}}} \nabla^2_{\vec{r}_N} + V_{\text{2b}} + 
V_{\text{3b}},
\end{eqnarray}
is designed to capture the
low-energy properties of $N$-body droplets.
The position vectors
of the bosons of mass $m_{\text{B}}$ are denoted
by $\vec{r}_j$ ($j=1,\cdots,N-1$) and the position
vector of the impurity of mass $m_{\text{X}}$ 
is denoted by $\vec{r}_N$.
The potential $V_{\text{2b}}$ accounts for the 
pairwise interactions between the bosons and the impurity,
$V_{\text{2b}}=\sum_{j=1}^{N-1} v_0 \exp[-r_{jN}^2/(2 r_0^2)]$,
where the depth $v_0$ ($v_0<0$) and the range $r_0$ are adjusted to 
reproduce the desired interspecies two-body scattering length $a_s$
and $r_{jk}$ is equal to $|\vec{r}_j-\vec{r}_k|$.
Motivated by our desire to explore the extension of Efimov's
BBX trimer study with large BX and vanishing BB
$s$-wave scattering 
lengths~\cite{BraatenReview,Efimov2,Efimov3,petrov,esry2,esry1,hammer2}, 
which has been realized
experimentally~\cite{Chicago,Heidelberg}, to the $N>3$ sector,
we neglect the interactions between the identical heavy bosons.

The potential $V_{\text{3b}}$ accounts for a repulsive
three-body force 
for each BBX triple,
$V_{\text{3b}}=\sum_{j<k}^{N-1} V_0 \exp[-
(r_{jk}^2+r_{jN}^2+r_{kN}^2)
/(2 R_0^2)]$~\cite{vonStecherJPB,commentthreebodyforce}. 
For diverging BX scattering length,
the height $V_0$ and range $R_0$ of the repulsive three-body interaction
are adjusted such that the 
lowest trimer state is much larger than $r_0$ and $R_0$, i.e.,
such that the wave function of the lowest trimer is
insensitive to the details of the model interactions
and accurately described by Efimov's
zero-range theory~\cite{comment4,supplement}.
Throughout, we use $R_0=\sqrt{8}r_0$.
Having fixed the parameters of the model Hamiltonian by
analyzing the properties of the three-body system,
the four- and higher-body sectors are explored and found to be
universal, i.e., the four- and higher-body observables are
found to be largely insensitive to the details of
the underlying potential model, provided the $N$-body ($N>3$) observables 
are expressed in terms of the corresponding three-body observables.
We emphasize that our model Hamiltonian
does not allow us to predict the three-body parameter, which
is expected to be determined by the long-range
van der Waals tail of the true atom-atom 
interactions~\cite{grimmPRL,schmidt,wangvdw1,wangvdw2,naidonvdw1,naidonvdw2}.
Rather, the model Hamiltonian allows us to predict
four- and higher-body properties relative to
the three-body properties. The underlying premise is that
the four- and higher-body sectors are fully determined by
the three-body sector.

To solve the time-independent Schr\"odinger equation for the
Hamiltonian given in Eq.~(\ref{eq_ham}),
we expand the eigenstates in the relative coordinates in
terms of explicitly correlated Gaussian basis 
functions~\cite{supplement,CGbook,CGRMP,debrajPRA}.
The resulting eigenenergies $E_N$ provide,
according to the Hylleraas-Undheim-MacDonald theorem,
variational
upper bounds to the energies of the ground and excited states of the 
system~\cite{supplement,CGbook,CGRMP}.
The states considered in this work have vanishing angular
momentum and positive parity. 
Since our implementation
provides access only to true bound states and not to resonance states,
we are limited
to treating $N$-body states that lie below the ground state 
of the $(N-1)$-body system, i.e.,
we have access, provided they exist, to $N$-body states
that are tied to the lowest Efimov trimer and not
to those that are tied to energetically higher-lying Efimov trimers.

To validate our approach, we consider the $N$ identical
boson system with infinitely large $s$-wave scattering 
length~\cite{supplement}.
We find $(E_4^{(1,1)}/E_3^{(1)})^{1/2}=2.127(5)$
and $(E_5^{(1,1)}/E_3^{(1)})^{1/2}=3.21(5)$, which agrees
well with the literature values of $2.147$~\cite{Deltuva} and 
$3.22(4)$~\cite{vonStecherJPB}.
For the four-body system, the discrepancy can be explained by
small finite-range corrections.
Moreover, our calculations confirm the existence of an
extremely weakly-bound excited tetramer~\cite{Platter,vonStecher,Deltuva}.

\begin{figure}
\vspace*{+.cm}
\includegraphics[angle=0,width=70mm]{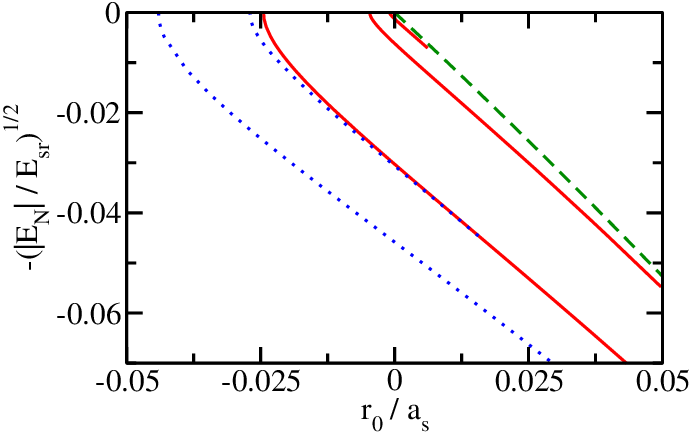}
\vspace*{0.5cm}
\caption{(Color online)
Efimov plot for the B$_{N-1}$X system ($N=3$ and $4$) with $\kappa=133/6$.
The dashed line shows the energy of the BX system.
The solid lines show the three lowest energies of the BBX system.
The dotted lines show the energies of the
two bound states of the BBBX system that are
tied to the lowest Efimov trimer.
The excited tetramer becomes unbound at 
$r_0/a_s \approx 0.02$.
The calculations are
performed for $V_0=3.2E_{\text{sr}}$,
where $E_{\text{sr}}$ is equal to $\hbar^2/(2 \mu r_0^2)$
and $\mu$ denotes the reduced mass, 
$\mu = m_{\text{B}} m_{\text{X}} /(m_{\text{B}}+ m_{\text{X}})$.}
\label{fig1}
\end{figure}

Figure~\ref{fig1} shows the extended Efimov plot for the
Cs$_{N-1}$Li system  with $N=3$ and 4~\cite{massratio}.
The energies of the dimer, trimer and tetramer states are shown
by dashed, solid, and dotted lines, respectively.
The energy ratios between consecutive trimers at unitarity 
are close to those predicted
by the universal zero-range theory
(see Table~\ref{tab1}). For the lowest two trimers, 
the ratio deviates from the universal value
by $0.8\%$, indicating that finite-range effects are
negligibly small near unitarity.
Non-universal finite-range corrections do, however, play a role 
when the trimers merge with the
three-atom
and atom-dimer thresholds.
The scattering length ratios where the
trimers hit the three-atom threshold are
found to be $a_{3,-}^{(2)}/a_{3,-}^{(1)} = 5.28(8)$
and $a_{3,-}^{(3)}/a_{3,-}^{(2)} = 4.95(8)$, 
which deviate by $8.5\%$ and $1.7\%$, respectively,
from the universal zero-range theory value of $4.865$.

For negative and sufficiently large positive interspecies 
scattering lengths, we find two tetramers 
that are bound with respect to the lowest
trimer. The energies of these tetramers ``trace'' the energy of the lowest
trimer.
At unitarity, we find $(E_4^{(1,1)}/E_3^{(1)})^{1/2}=1.510(5)$
and $(E_4^{(1,2)}/E_3^{(1)})^{1/2}=1.010(5)$. These values are 
expected to be fairly close to what the universal zero-range
theory would yield.
At $a_s \approx 2.6(4) a_{\text{td}}^{(1)}$, where 
$a_{\text{td}}^{(1)}$ denotes the scattering length where the
lowest trimer energy is equal to that of two dimers,
the energy of the excited tetramer is equal to that of the lowest 
trimer, indicating that the excited tetramer 
becomes unbound at this scattering length~\cite{comment1}.

Qualitatively, the energy spectrum shown in Fig.~\ref{fig1}
is  similar to that for the $N$ identical boson system,
which supports two universal tetramers for 
$1/a_s \le 1/[13.75(5) a_{\text{td}}^{(n)}]$~\cite{Deltuva}.
In that system, it has been shown that the universal tetramers
are not only attached to the lowest Efimov trimer but to
each Efimov trimer (for the excited Efimov trimers, the ``attached''
four-body states correspond to resonance states~\cite{Deltuva}). 
We conjecture that this is also true for the
B$_3$X system, i.e., we conjecture that there exist two
tetramers with energies $E_{4}^{(n,1)}$ and $E_4^{(n,2)}$
that are universally tied to the $n$th Efimov trimer for 
$1/a_s$ smaller than a critical inverse scattering length.
On the positive
scattering length side, we restricted our four-body calculations
to fairly large $a_s$. As $a_s$
decreases, the spectrum has been predicted to 
contain additional four-body states~\cite{Wang}, 
which can be thought of
as corresponding to  Efimov trimer states 
consisting of a dimer and two atoms.

We find that the scattering lengths where the
four-body states merge with the four-atom threshold are 
given by $a_{4,-}^{(1,1)} \approx 0.55 a_{3,-}^{(1)}$ 
and $a_{4,-}^{(1,2)} \approx 0.91 a_{3,-}^{(1)}$
for the ground and excited tetramers, respectively.
Due to finite-range effects,
these ratios are expected to differ somewhat from those values that the 
universal zero-range theory would predict.
The fact that
the excited tetramer is very weakly bound with respect to the trimer 
implies that
the scattering length $a_{3,-}^{(1)}$
at which the trimer is in resonance with the
three-atom threshold and the scattering length $a_{4,-}^{(1,2)}$
at which the excited tetramer is in resonance 
with the four-atom threshold are quite close.
Taking the value of $a_{3,-}^{(1)}=-337(9)a_0$ [$-320(10)a_0$]
from the Chicago~\cite{Chicago} [Heidelberg~\cite{Heidelberg}] experiment, 
this yields
$a_{4,-}^{(1,1)} \approx -187 a_0$ [$-178 a_0$] and 
$a_{4,-}^{(1,2)} \approx -305 a_0$ [$-290 a_0$].
Our results suggest that the analysis of the experimental data 
could be impacted by the existence of the excited tetramer discussed
in the present work.
Future work should 
disentangle the zero- and finite-range
effects, and possibly build van der Waals universality
into the model Hamiltonian.
Moreover, finite temperature effects need to be investigated carefully.

\begin{widetext}

\begin{table}
\caption{Energies of the B$_{N-1}$X system 
with infinitely large interspecies $s$-wave scattering length
for various mass ratios.
The second column reports the binding momentum of the lowest trimer state in
units of the binding momentum of the
short-range energy scale $E_{\text{sr}}$.
Columns 3-7 report ratios of binding momenta for the systems
with $N=3-6$. For columns 3-6,
the energies were extrapolated to the 
$V_{0} \rightarrow \infty$ limit. 
For columns 7-8, we used $V_0=3.2 E_{\text{sr}}$. 
The symbol ``---'' indicates that no such bound state was found.
In the cases where no entry is given, calculations were not
performed.
For comparison,
the last column reports the scaling factor $\lambda$ calculated
from Efimov's zero-range theory.
}
\label{tab1}
\begin{tabular}{c|ccccccc|c}
$\kappa$ & 
$\sqrt{E_3^{(1)}/E_{\text{sr}}}$ & $\sqrt{{E_3^{(2)}}/{E_3^{(1)}}}$ &
$\sqrt{{E_3^{(3)}}/{E_3^{(2)}}}$ &
$\sqrt{{E_{4}^{(1,1)}}/{E_3^{(1)}}}$ &
$\sqrt{{E_{4}^{(1,2)}}/{E_3^{(1)}}}$ &
$\sqrt{{E_{5}^{(1,1)}}/{E_3^{(1)}}}$ &
$\sqrt{{E_{6}^{(1,1)}}/{E_3^{(1)}}}$ &
$\lambda=e^{-\pi/s_0}$ 
\\ \hline
$8$     & $0.012$ & $12.510(5)$ && $1.647(5)$ & --- & $2.06(4)$ && $12.4878$ \\
$12$    & $0.017$ & $8.158(5)$ && $1.58(1)$ & --- & $1.94(4)$ && $8.1305$ \\
$16$    & $0.021$ & $6.313(5)$ && $1.544(5)$ & $1.002(1)$ & $1.88(4)$ && $6.2804$ \\
$133/6$ & $0.024$ & $4.904(5)$ & $4.867(2)$ & $1.510(5)$ & $1.010(5)$ & $1.82(4)$ & $2.03(10)$ & $4.8651$ \\
$30$    & $0.028$ & $3.998(3)$ & $3.958(3)$ & $1.488(5)$ & $1.026(5)$ & $1.78(4)$ & $1.95(10)$ & $3.9553$ \\
$40$    & $0.031$ & $3.372(3)$ & $3.330(2)$ & $1.471(5)$ & $1.046(8)$ & $1.75(4)$ && $3.3249$ \\
$50$    & $0.033$ & $2.996(5)$ & $2.952(4)$ & $1.461(5)$ & $1.067(8)$ & $1.73(4)$ && $2.9470$ \\
\end{tabular}
\end{table}
\end{widetext}

We now discuss extensions of Fig.~\ref{fig1}
to other mass ratios $\kappa$ and larger $N$.
Our results for infinitely large $a_s$
are summarized in Table~\ref{fig1}.
The lowest tetramer becomes less strongly bound
with respect to the lowest
Efimov trimer with increasing mass ratio and appears to approach
a constant for large $\kappa$. 
The ratio $(E_4^{(1,2)}/E_3^{(1)})^{1/2}$ at unitarity increases from
$1.002(1)$ for $\kappa=16$ to $1.067(8)$ for $\kappa=50$.
Our results disagree with
a recent study that reported that BBBX systems with mass ratios
$\kappa=30$ and $50$
support a single tetramer state tied to each Efimov trimer~\cite{Wang}.
The reason for this disagreement is not clear.
We find that the excited tetramer 
appears at $\kappa \approx 13$.
For $\kappa=12$ and $8$,
we find an excited tetramer that is bound
relative to the lowest trimer for negative scattering lengths away
from unitarity but not at unitarity. 
This shows that the excited tetramer ceases
to exist
at different $a_s$ for $\kappa=8$ to $133/6$.
We did not investigate what happens to the excited tetramer
for $\kappa=30-50$ on the positive scattering length side.
For $\kappa = 50$ and infinitely large interspecies
scattering length, we 
searched for a second excited tetramer with energy $E_4^{(1,3)}$ that
is bound with respect to the lowest Efimov trimer 
but did not find one.
The energies of the
lowest $N=5$ and 6 states (see columns 7-8 of Table~\ref{tab1})
behave similar to the
energy of the lowest tetramer,
i.e., the binding of the lowest pentamer 
relative to the lowest tetramer 
and the binding of the lowest $N=6$ state relative to the lowest
pentamer decrease with increasing $\kappa$.
It would be interesting to extend the calculations presented
in this paper to larger $N$.

\begin{figure}
\vspace*{+.5cm}
\includegraphics[angle=0,width=65mm]{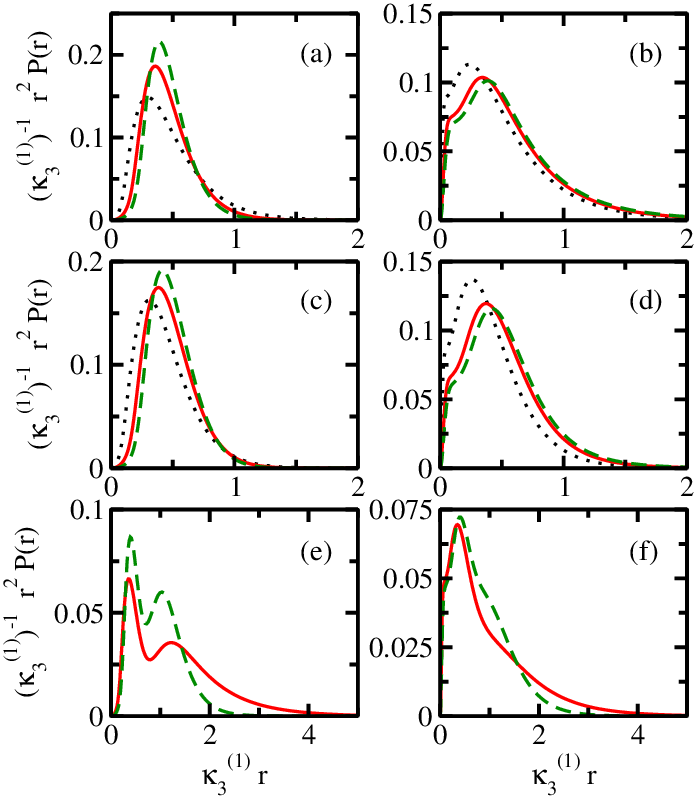}
\vspace*{0.cm}
\caption{(Color online)
Pair distribution functions
for infinitely large BX and vanishing BB scattering lengths
for (a) and (b) the lowest trimer,
(c) and (d) the lowest tetramer, and
(e) and (f) the first excited tetramer.
Panels (a), (c) and (e)
show the scaled pair distribution functions
$r^2 P_{\text{BB}}(r)$ while
panels (b), (d) and (f)
show the scaled pair distribution functions
$r^2 P_{\text{BX}}(r)$.
Dotted, solid, and dashed lines are for $\kappa=8$,
$133/6$ and $40$, respectively.
The calculations are performed for $V_{3\text{b}}=3.2 E_{\text{sr}}$.
}
\label{fig_pair}
\end{figure}

Figure~\ref{fig_pair}
shows the pair distribution functions
for the B$_{N-1}$X systems 
($N=3$ and $4$) with $\kappa=8$ (dotted line), 
$133/6$ (solid line) and $40$
(dashed line) for infinitely large BX and vanishing BB
scattering lengths.
The scaled pair distribution function
$4 \pi r^2 P_{\text{BB}}(r)$ (left column of Fig.~\ref{fig_pair})
tells one the likelihood to find two identical bosons at
a distance $r$ from each other while
the scaled pair distribution function
$4 \pi r^2 P_{\text{BX}}(r)$ (right column of Fig.~\ref{fig_pair})
tells one the likelihood to find a B atom 
at
a distance $r$ from the X atom.
To facilitate the comparison between systems with different mass ratios,
the lengths in Fig.~\ref{fig_pair} are scaled 
by the binding momentum $\kappa_3^{(1)}$, where
$\hbar \kappa_3^{(1)}=(2 \mu E_3^{(1)})^{1/2}$~\cite{comment2}.

Figures~\ref{fig_pair}(a) and \ref{fig_pair}(b) 
show $r^2 P_{\text{BB}}(r)$ and $r^2 P_{\text{BX}}(r)$
for the lowest Efimov trimer.
Two characteristics
are evident. First, $r^2 P_{\text{BX}}$
is larger at small $r$ than $r^2 P_{\text{BB}}$.
This is not surprising, as the
B atoms do not interact and are held together through the 
light impurity.
Second, the B atoms become slightly more localized with increasing mass ratio
$\kappa$, i.e., the BB pair distribution function becomes 
narrower with increasing $\kappa$.
Figures~\ref{fig_pair}(c)-\ref{fig_pair}(f)
show the scaled pair distribution functions for the BBBX system.
The scaled pair distribution
functions for the lowest tetramer 
[Figs.~\ref{fig_pair}(c)-\ref{fig_pair}(d)]
behave similarly to those for the lowest trimer.
For the excited tetramer
[Figs.~\ref{fig_pair}(e)-\ref{fig_pair}(f)], the scaled pair distribution
functions exhibit a double-peak structure (BB distance)
or ``shoulder'' at large distances (BX distance),
indicating that the excited tetramer can be roughly thought of,
like the excited tetramer in the four identical boson 
system~\cite{vonStecherPRL}, as a 
trimer with a 
fourth atom ``tagged on'' (i.e., a ``3+1 state'').

In summary,
this paper presented results for the extended Efimov scenario
for heteronuclear B$_{N-1}$X mixtures.
It was found that the number of universal four-body bound states
that are tied to the Efimov trimers depends on the mass ratio
and scattering length.
Structural properties of the four-body system were analyzed and
extensions to the five- and six-body sector were presented.
The results presented constitute an important contribution to
the understanding of universal low-energy phenomena
across the fields of atomic, nuclear and particle
physics.
Our calculations present the first comprehensive study of
the extension of the generalized Efimov scenario to
heteronuclear mixtures
and are directly relevant to 
on-going cold atom experiments on ultracold Cs-Li mixtures.
Concretely, an estimate of the four-atom resonance positions
was given.

{\bf{Acknowledgments:}}
DB acknowledges 
Hans-Werner Hammer and Shina Tan
for insightful discussions, and
Yujun Wang for
sharing unpublished data related to Ref.~\cite{Wang}.
Support by the National
Science Foundation (NSF) through Grant No.
PHY-1205443
is gratefully acknowledged.
DB also acknowledges support from the Institute for Nuclear Theory
during the program  INT-14-1,
``Universality in Few-Body Systems: Theoretical
Challenges and New Directions''.


\begin{thebibliography}{100}
\bibitem{energyscale}
Throughout this work, an energy of zero corresponds to 
two infinitely far separated atoms.
The three- and four-body energies are measured relative to this
``two-body zero-energy threshold.''
\bibitem{footnote2vs3}
The Efimov regime for three-body systems 
can be realized when two of three two-body $s$-wave 
scattering lengths are large in magnitude~\cite{BraatenReview}.
\bibitem{commentthreebodyforce}
Note that the three-body potential employed in this paper 
is parameterized in terms of the sum of the square of the 
three interparticle distances and not in terms of the
hyperradius of the three-body subsystem.
For the equal-mass system, the two parameterizations are, 
except for a scaling factor, equivalent~\cite{vonStecherJPB}.
\bibitem{supplement}
See to-be-inserted-by-the-editor for
details on the comparison with the zero-range theory results, and
for details on the basis set expansion approach and 
other computational
aspects.
\bibitem{massratio}
The mass ratio for Cs-Li is $22.095$, which is $0.3$\% smaller than the
value $133/6\approx 22.167$ used in the present work.
The zero-range theory yields $\lambda=4.877$ and $4.865$
for $\kappa=22.095$ and $22.167$, respectively.
The difference between these $\lambda$ values is $0.25$\%.
For the purpose of the present paper, the difference in the mass
ratios is insignificant.
\bibitem{comment1}
The scattering length at which the energy of the excited
tetramer crosses the energy of the lowest trimer is expected
to be sensitive to finite-range effects, i.e., a
precise determination of the critical scattering length value requires
precise knowledge of the energy difference $E_4^{(n,2)}-E_3^{(n)}$.
\bibitem{comment2}
Note that
the use of the two-body reduced mass $\mu$ is, to some extent,
arbitrary; one could alternatively use a three-body reduced mass.
\end{thebibliography}
\end{document}


\title{Supplemental material for ``Generalized Efimov scenario for heavy-light mixtures''}

\author{D. Blume and Yangqian Yan}
\affiliation{Department of Physics and Astronomy,
Washington State University,
  Pullman, Washington 99164-2814, USA}

\date{\today}

\begin{abstract}
The notation employed in this supplemental
material follows that introduced in the main text.
\end{abstract}


\maketitle

\section{Basis set expansion approach}

To solve the time-independent 
Schr\"odinger equation for  the
Hamiltonian $H$ given in Eq.~(1) 
of the main text,
we employ an explicitly correlated Gaussian basis 
set~\cite{CGbook,CGRMP}.
The eigenfunctions $\psi_{\beta}$
of the Hamiltonian $H$ 
are expanded in terms of the basis functions $\phi_l^{(\beta)}$,
\begin{eqnarray}
\label{eq_expansion}
\psi_{\beta} = \sum_{l=1}^{N_b}  c_l^{(\beta)} \phi_l^{(\beta)},
\end{eqnarray}
where each of the basis functions $\phi_l^{(\beta)}$,
\begin{eqnarray}
\label{eq_ecg}
\phi_l^{(\beta)}={\cal{S}} \exp \left[
- \frac{1}{2} 
\sum_{j=1}^{N-1} \sum_{k>j}^N 
\left(\frac{r_{jk}}{\alpha_{jk}^{(l)}}
\right)^2
\right],
\end{eqnarray}
depends on $N(N-1)/2$ independent non-linear variational
parameters
$\alpha_{jk}^{(l)}$ that are optimized semi-stochastically.
For notational simplicity, the dependence of the $\alpha_{jk}^{(l)}$'s
on the state index $\beta$ is not indicated explicitly
in Eq.~(\ref{eq_ecg}).
${\cal{S}}$ denotes a symmetrizer that ensures that 
the basis function $\phi_l^{(\beta)}$ is symmetric under the
exchange of any two identical bosons.
The $c_l^{(\beta)}$ denote linear variational or expansion parameters
that are determined by solving the generalized
eigenvalue problem
$\underline{H} \vec{c}^{(\beta)} = 
E^{\text{var}}_{\beta} \underline{O} \vec{c}^{(\beta)}$,
where $\underline{H}$ and $\underline{O}$ denote the Hamiltonian and
(non-diagonal) overlap matrices, respectively.
The vector $\vec{c}^{(\beta)}$ contains the coefficients
$c_1^{(\beta)},\cdots,c_{N_b}^{(\beta)}$, where $N_b$ denotes the size of the
matrix (or equivalently, the size of the basis set).
According to the variational principle,
the energies
$E^{\text{var}}_{\beta}$ are upper bounds to the exact eigenenergies
$E_{\beta}$.
Assuming that $E_1^{\text{var}} \le E_2^{\text{var}} \le \cdots \le E_{N_b}^{\text{var}}$, one has
$E_1 \le E^{\text{var}}_1, E_2 \le E^{\text{var}}_2,\cdots$.
The matrix elements $H_{ll'}$ and $O_{ll'}$ have closed analytical expressions
and 
the generalized eigenvalue problem
is solved using one of ARPACK's eigenvalue solvers.

The superscript ``$(\beta)$''
on the right hand side of Eq.~(\ref{eq_expansion})
indicates that the basis set is constructed for the
$\beta$th eigenstate $\psi_{\beta}$.
While one could construct a single basis set that provides
a good description of the lowest few eigenstates,
our work takes advantage of the fact that the basis set can be optimized 
separately for each eigenstate.
For the BBX system, e.g., the two energetically
lowest-lying states differ in size by the scaling factor $\lambda$.
This implies that the variational parameters $\alpha_{jk}^{(l)}$
that yield an efficient description of
the ground state (the state with $\beta=1$) and of the
first excited state (the state with $\beta=2$)
are very different.
Another key point of the basis set expansion approach is that the
basis set can be systematically improved.
Our three-body energies are, except very close to the three-atom
break-up threshold, converged to $0.1$\% or better.
Our four-body energies are converged to 1\% or better.
For the B$_3$X system 
at unitarity with $\kappa=133/6$, e.g., we clearly
see that the energy of the first excited four-body state 
lies below that of the lowest BBX state.

\section{Benchmarking our approach: $N$ identical bosons}

To validate our approach, we consider $N$ identical bosons
of mass $m_{\text{B}}$ with infinitely large
$s$-wave scattering length
$a_s$ described by the Hamiltonian $H_{\text{B}}$,
\begin{eqnarray}
\label{eq_hambosons}
H_{\text{B}} = \sum_{j=1}^N -\frac{\hbar^2}{2m_{\text{B}}} \nabla^2_{\vec{r}_j} +
V_{2\text{b}}+V_{3\text{b}}.
\end{eqnarray}
The potential
$V_{2\text{b}}$ accounts for the interactions between all
$N(N-1)/2$ pairs,
\begin{eqnarray}
V_{2\text{b}}=\sum_{j=1}^{N-1} \sum_{k>j}^N v_0 
\exp \left(- \frac{r_{jk}^2}{2r_0^2}
\right),
\end{eqnarray}
where $v_0$ and $r_0$ denote the depth and range of the attractive 
two-body Gaussian. The depth and range are adjusted such that
the free-space two-body system
supports one zero-energy bound state.
The potential $V_{3\text{b}}$ accounts for the interactions
between 
all $N(N-1)(N-2)/6$ triples,
\begin{eqnarray}
V_{3\text{b}}= 
\sum_{j=1}^{N-2} \sum_{k>j}^{N-1} \sum_{l>k}^N
V_0 \exp \left(-\frac{r_{jk}^2+r_{kl}^2+r_{lj}^2}{2R_0^2}
\right),
\end{eqnarray}
where
$V_0$ and $R_0$ denote the depth and range of the repulsive
three-body Gaussian.
Our calculations use $R_0=\sqrt{8} r_0$. As discussed in 
Ref.~\cite{vonStecherJPB},
the repulsive three-body potential
serves to eliminate deeply-bound non-universal 
states in the $N=3$ sector.
In essence, the three-body potential ``cuts off'' the short-range
portion of the effective hyperradial potential curve that is governed by the
two-body effective range and deviates from the effective three-body
hyperradial Efimov potential curve.
Using this model,
the ratio of the binding momenta of the two energetically
lowest-lying three-body states is
$(E_3^{(1)}/E_3^{(2)})^{1/2}=22.99$ for $V_0=0$,
takes a minimum value of $(E_3^{(1)}/E_3^{(2)})^{1/2}=21.48$ 
for $V_0 \approx 0.3 E_{\text{sr}}$,
and approaches 
$(E_3^{(1)}/E_3^{(2)})^{1/2}=22.71$ 
as $V_0 \rightarrow \infty$. 
For $V_0 \ge E_{\text{sr}}$, the ratio
$E_3^{(1)}/E_3^{(2)}$ lies within 0.08\% of the universal zero-range value 
of $22.694$.
The small difference of the binding momentum ratios
for the finite-range Hamiltonian
$H_{\text{B}}$ and for the zero-range model can be attributed to the fact 
that both $V_{2\text{b}}$ and $V_{3\text{b}}$ have a finite range.

The four identical boson system with infinitely large $s$-wave 
scattering length has been benchmarked most precisely by 
Deltuva~\cite{Deltuva}
using a momentum space representation
that allows for the treatment of bound and resonance states.
It was shown that the four-body energies $E_4^{(n,1)}$ 
and $E_4^{(n,2)}$
approach the ratios
$(E_4^{(n,1)}/E_3^{(n)})^{1/2}=2.147$ and 
$(E_4^{(n,2)}/ E_3^{(n)})^{1/2}=1.0011$, respectively, for sufficiently 
large $n$~\cite{Deltuva}.
For the Hamiltonian given in Eq.~(\ref{eq_hambosons}) 
with $V_0=4.8 E_{\text{sr}}$,
we find (see also the main text)
that the binding momentum ratio of the lowest four-body state
and the lowest three-body state 
is 
$(E_4^{(1,1)}/E_3^{(1)})^{1/2}=2.127(5)$.
For this $V_0$,
the lowest three-body energy is 
$-2.64  \times 10^{-4} E_{\text{sr}}$, 
i.e., the
three-body system is
large compared to both
$r_0$ and $R_0$ and thus, to a good
approximation, independent of $r_0$ and $R_0$.
We also find a weakly-bound excited
four-body state with the binding momentum ratio
$(E_4^{(1,2)}/E_3^{(1)})^{1/2} \ge 1.0004$
that is tied to the lowest
three-body state.
Although the variational principle does not apply to energy
ratios, we can assign the ``$\ge$'' sign since the energy of the lowest
three-body state has a significantly smaller basis set error than
the energy of the excited four-body state.

\section{Heavy-light $(2,1)$ system at unitarity}

\begin{figure}
\vspace*{+.cm}
\includegraphics[angle=0,width=70mm]{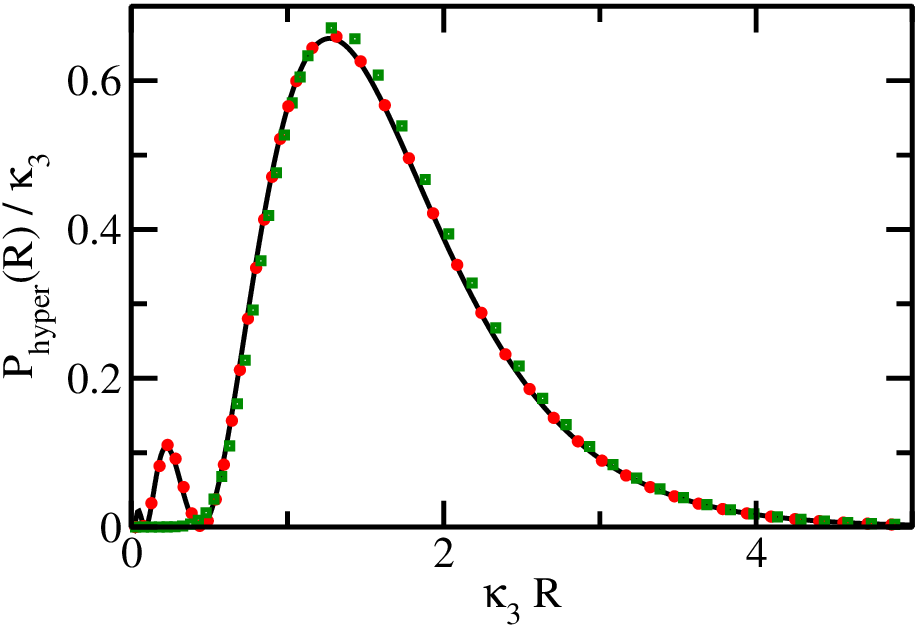}
\vspace*{0.5cm}
\caption{(Color online)
Hyperradial density $P_{\text{hyper}}(R)$ 
for the B$_2$X system with
infinitely large 
interspecies $s$-wave scattering length and
$\kappa=133/6$.
The squares and circles show the hyperradial densities
for the ground and first excited state of the
finite-range model Hamiltonian
with $V_0=3.2 E_{\text{sr}}$ and $R_0=\sqrt{8}r_0$.
The solid line shows the hyperradial density for the
zero-range model Hamiltonian.
For all three curves, dimensionless units are used (see 
the text for details).}
\label{fig1}
\end{figure}

To validate our calculations for the $(2,1)$ system
with unequal masses,
we consider the case where the interspecies 
$s$-wave scattering length is infinitely large and
the intraspecies two-body potential is set to zero.
In the limit of zero-range interactions, the
hyperangular and hyperradial degrees of freedom
separate, and 
the hyperradial density $P_{\text{hyper}}(R)$
can be calculated analytically~\cite{BraatenReview}
(see the solid line in
Fig.~\ref{fig1} for $\kappa=133/6$).
Here, the hyperradius $R$ is defined through
\begin{eqnarray}
\label{eq_rhyper}
\mu R^2= \sum_{j=1}^{2} m_{\text{B}} (\vec{r}_j - \vec{r}_{\text{cm}})^2
+ m_{\text{X}} (\vec{r}_3 - \vec{r}_{\text{cm}})^2,
\end{eqnarray}
where $\vec{r}_{\text{cm}}$ denotes the 
center-of-mass vector of the B$_2$X system.
For comparison,
squares and circles show the
hyperradial densities 
for the energetically lowest-lying and 
second lowest-lying states 
obtained from our basis set expansion calculations for 
the Hamiltonian $H$ [see Eq.~(1) of the main text].
To make this figure,
the lengths have been scaled by the 
three-body parameter
$\kappa_3$.
For the circles and the solid line,
$\kappa_3$ is defined through
$\hbar^2 \kappa_3^2/(2\mu)=|E_3^{(2)}|$,
where $E_3^{(2)}$ is the
energy of the first excited state of 
the finite-range model Hamiltonian.
For the squares, we define $\kappa_3$ 
through $\hbar^2 \kappa_3^2/(2\mu) = \lambda^2 |E_3^{(2)}|$,
where $\lambda$ is obtained by solving the 
hyperangular portion of the zero-range
model Hamiltonian.
As shown in Table~I of the main text, this zero-range 
scaling factor is very close to
the scaling factor obtained from the spacing between the two
lowest three-body energies of the finite-range model
Hamiltonian.
The good agreement
for
$\kappa_3R \gtrsim 0.5$ 
between the hyperradial density for the 
zero-range model and the hyperradial densities 
of the
finite-range Hamiltonian with two-
and three-body interactions demonstrates that the model
Hamiltonian employed in our work captures
the Efimov
physics in the three-body sector accurately.

\begin{figure}
\vspace*{+.cm}
\includegraphics[angle=0,width=70mm]{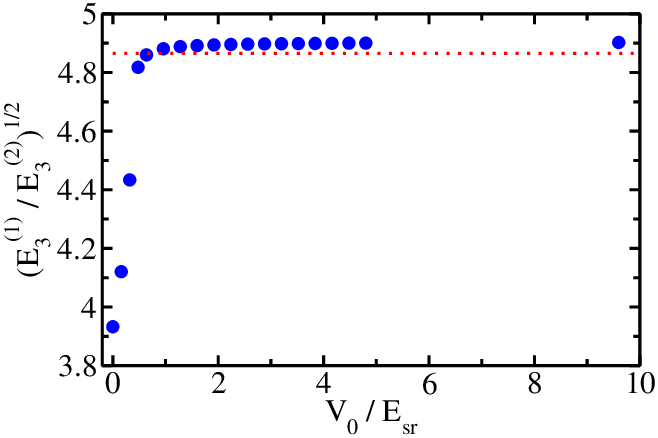}
\vspace*{0.5cm}
\caption{(Color online)
Symbols show 
the binding momentum ratio $(E_3^{(1)}/E_3^{(2)})^{1/2}$ as a function of $V_0$
for the B$_2$X system with infinitely large
interspecies $s$-wave scattering
length and $\kappa=133/6$.
For these calculations, we used $R_0=\sqrt{8}r_0$.
The dotted line shows the binding momentum ratio for the
zero-range model.}
\label{fig2}
\end{figure}

Varying
$V_0$ changes the three-body parameter.
The scaled hyperradial densities and energy ratios,
however,
are, to a good approximation, unchanged for $V_0 \gtrsim E_{\text{sr}}$
and
agree well with those for the zero-range model.
As an example,
Fig.~\ref{fig2}
shows the binding momentum ratio
$(E_3^{(1)}/E_3^{(2)})^{1/2}$ as a function of $V_0$ for infinitely
large interspecies $s$-wave scattering length
and $\kappa=133/6$.
As can be seen, the binding momentum ratio is approximately
independent 
of $V_0$ for $V_0 \gtrsim  E_{\text{sr}}$.
In the large $V_0$ limit,
the binding momentum ratio is close but not identical to 
the binding momentum 
ratio predicted by the zero-range theory.
The small deviation can be attributed to the weak breaking of the discrete scale invariance
of the model Hamiltonian by the finite-range two- and three-body
interactions.

\section{Heavy-light $(3,1)$ system}

\begin{figure}
\vspace*{+1.cm}
\includegraphics[angle=0,width=70mm]{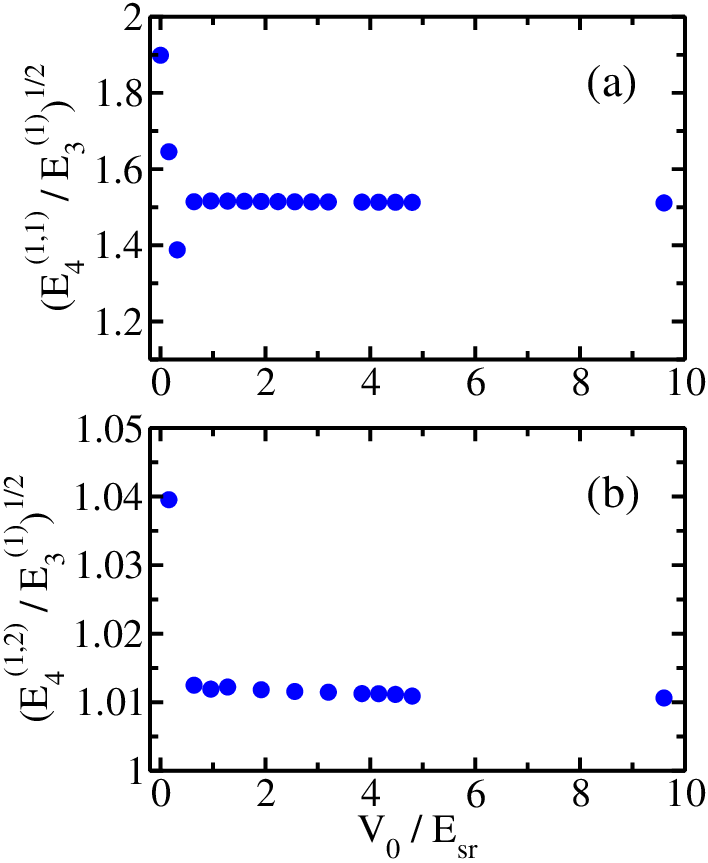}
\vspace*{0.5cm}
\caption{(Color online)
Symbols show the
binding momentum ratios (a) $(E_4^{(1,1)}/E_3^{(1)})^{1/2}$ and
(b) $(E_4^{(1,2)}/E_3^{(1)})^{1/2}$ 
as a function of $V_0$
for the heavy-light system with infinitely large 
interspecies $s$-wave scattering
length and $\kappa=133/6$.
For these calculations, we used $R_0=\sqrt{8}r_0$.}
\label{fig3}
\end{figure}

Symbols in Figs.~\ref{fig3}(a) 
and \ref{fig3}(b) show the binding momentum ratios
$(E_4^{(1,1)}/E_3^{(1)})^{1/2}$ and
$(E_4^{(1,2)}/E_3^{(1)})^{1/2}$,
respectively, as a function of $V_0$
for infinitely large interspecies $s$-wave scattering length
$a_s$, vanishing intraspecies interactions and $\kappa=133/6$.
The binding momentum 
ratios are approximately independent of $V_0$ for 
$V_0 \gtrsim E_{\text{sr}}$.
This suggests that the properties of
the B$_3$X system are, to a good approximation, determined by 
the two-body $s$-wave scattering length $a_s$ and the
three-body parameter $\kappa_3$.

To further test the robustness of our results
against changes of the parameters in the model Hamiltonian,
we considered a 1.5 times larger range of the 
repulsive three-body  potential while keeping the two-body 
range $r_0$ unchanged.
The resulting change in the observables was found to be 
quite small.
In addition, we varied the 
functional form of $V_{3\text{b}}$, i.e., we considered a repulsive
three-body potential that is parameterized in terms
of the hyperradii of the B$_2$X subsystems 
as opposed to the sum of the squares of the interparticle distances.
The key difference 
between this alternative parametrization and the parametrization 
employed earlier
is that this alternative three-body potential
does, for the B$_2$X system, not depend on the hyperangles;
note, however, that this alternative $V_{3\text{b}}$ does depend on the 
hyperangles for $N \ge 4$.
For this three-body potential, the four-body states
are bound a bit more weakly relative to 
the lowest three-body state than for the
three-body potential used in the main text.
At unitarity, we obtain $(E_4^{(1,1)}/E_3^{(1)})^{1/2}=1.47(2)$
and $(E_4^{(1,2)}/E_3^{(1)})^{1/2} \ge 1.004$.
For the scattering lengths at which the 
four-body system becomes unbound, we find
$a_{4,-}^{(1,1)} \approx 0.57 a_{3,-}^{(1)}$
and 
$a_{4,-}^{(1,2)} \approx 0.92 a_{3,-}^{(1)}$.
For comparison, the corresponding values reported in the main text are
$0.55$ and $0.91$, respectively.
The mass ratio at which the excited four-body state
ceases to exist at unitarity changes from approximately $13$ 
(this is the value reported in the main text) to $17$
for the alternative three-body potential.
These calculations suggest that the generalized Efimov
scenario
discussed in the main text is fairly robust with respect to changes
in the model Hamiltonian.